\documentclass[a4paper,12pt]{cernart}
\usepackage[dvips]{graphicx}
\usepackage{color}

\begin {document}

\newcommand{\epr}{\varepsilon^\prime\!/\varepsilon}
\newcommand{\repr}{\mathrm{Re}\,(\epr)}
\newcommand{\ko}{K^0}
\newcommand{\kobar}{\bar{K^0}}
\newcommand{\kl}{K_L}
\newcommand{\ks}{K_S}
\newcommand{\pipin}{\pi^0\pi^0}
\newcommand{\pipic}{\pi^+\pi^-}
\newcommand{\kethree}{K_{e3}}
\newcommand{\kmuthree}{K_{\mu3}}
\newcommand{\ptprime}{p_t{}^\prime}
\newcommand{\asp}{\mathcal{A}}
\definecolor{shade}{gray}{0.6}
\newcommand{\proposeout}[2]{}

\begin{titlepage}
\docnum{CERN--EP/99--114}
\date{18 August 1999} 
\vspace{1.2cm}
\title{A NEW MEASUREMENT OF DIRECT CP VIOLATION \\
IN TWO PION DECAYS OF THE NEUTRAL KAON}
\vspace{1.0cm}
\begin{Authlist}
The NA48 Collaboration 
\end{Authlist}
\vspace{1.0cm}
\begin{abstract}\noindent
The NA48 experiment at CERN has performed a new measurement
of direct CP violation, based on data taken in 1997 by
simultaneously collecting  $\kl$ and $\ks$ decays
into $\pipin$ and $\pipic$. The result for the
CP violating parameter $\repr$ is
$(18.5 \pm 4.5 \mathrm{(stat)} \pm 5.8 \mathrm{(syst)})\times10^{-4}$.
\end{abstract}
\vspace{10cm}
\submitted{(Submitted to Physics Letters B)}
\end{titlepage}
\newpage
\thispagestyle{empty}
\newpage
\thispagestyle{empty}
\begin{center}
\small{
V.~Fanti, A.~Lai, D.~Marras, 
L.~Musa\footnote{Present address: CERN, CH-1211 Geneva 23, Switzerland} \\
{\small \em Dipartimento di Fisica dell'Universit\`a e Sezione
    dell'INFN di Cagliari, I-09100 Cagliari, Italy} \\[0.35cm]
A.J.~Bevan, T.J.~Gershon,
B.~Hay,
R.W.~Moore,\footnote{Present address: Physics-Astronomy Building,
Michigan State University, East Lansing, MI  48824 USA} 
K.N.~Moore,$^{2)}$ D.J.~Munday,
M.D.~Needham,\footnote{Present address: NIKHEF, PO Box 41882, 
1009 DB Amsterdam, The Netherlands} 
M.A.~Parker, S.F.~Takach,\footnote{Present address: Department of Physics, 
Wayne State University, Detroit, MI 48201, USA}
T.O.~White, S.A.~Wotton \\
{\small \em Cavendish Laboratory, University of Cambridge, 
    Cambridge, CB3 0HE, UK\/}\footnote{Funded by the UK
    Particle Physics and Astronomy Research Council} \\[0.35cm]
G.~Barr,
H.~Bl\"umer,\footnote{Present address: Universit\"at Karlsruhe (TH), 
Fakult\"at f\"ur Physik, and Forschungszentrum
Karlsruhe GmbH, Institut f\"ur Kernphysik, D-76128 Karlsruhe}
G.~Bocquet,
J.~Bremer, 
A.~Ceccucci, 
J.~Cogan,\footnote{Present address: DSM/DAPNIA - CEA Saclay, F-91191 Gif-sur-Yvette, France}
D.~Cundy, N.~Doble, W.~Funk, L.~Gatignon, 
A.~Gianoli,\footnote{Present address: Dipartimento di Fisica dell'Universit\`a e Sezione dell'INFN di Ferrara, I-44100 Ferrara, Italy}
A.~Gonidec, G.~Govi, P.~Grafstr\"om, G.~Kesseler, W.~Kubischta, 
A.~Lacourt, S.~Luitz,\footnote{Present address: SLAC, Stanford, CA 94309, USA}
J.P.~Matheys, A.~Norton, S.~Palestini,
B.~Panzer-Steindel, 
B.~Peyaud,$^{7)}$ 
D.~Schinzel, H.~Taureg, 
M.~Velasco, 
O.~Vossnack, H.~Wahl, G.~Wirrer \\
{\small \em CERN, CH-1211 Geneva 23, Switzerland} \\[0.35cm]
A.~Gaponenko,
V.~Kekelidze, D.~Madigojine,
A.~Mestvirishvili,\footnote{Present address: Dipartimento di Fisica 
dell'Universit\`a e Sezione dell'INFN di Perugia, I-06100 Perugia, Italy}
Yu.~Potrebenikov, G.~Tatishvili,
A.~Tkatchev, A.~Zinchenko \\ 
{\small \em Joint Institute for Nuclear Research, Dubna, Russian
    Federation}\\[0.35cm]
L.~Bertolotto,$^{1)}$ O.~Boyle, I.G.~Knowles, V.J.~Martin, 
H.L.C.~Parsons, K.J.~Peach, C.~Talamonti\footnote{Present address:
Dipartimento di Fisiopatologia Clinica dell'Universit\`a di Firenze, 
I-50134 Firenze, Italy}   \\
{\small \em Department of Physics and Astronomy, University of
    Edinburgh, JCMB King's Buildings, Mayfield Road, Edinburgh,
    EH9 3JZ, UK\/}$^{5)}$ \\[0.35cm]
M.~Contalbrigo, P.~Dalpiaz, J.~Duclos, 
A.~Formica,$^{7)}$
P.L.~Frabetti,\footnote{Permanent address: Dipartimento di 
Fisica dell'Universit\`a e Sezione dell'INFN di Bologna, 
I-40126 Bologna, Italy}
M.~Martini, F.~Petrucci, M.~Savri\'e  \\
{\small \em Dipartimento di Fisica dell'Universit\`a e Sezione
    dell'INFN di Ferrara, I-44100 Ferrara, Italy} \\[0.35cm]
A.~Bizzeti,\footnote{Also at Dipartimento di Fisica 
dell'Universit\`a di Modena, I-41100 Modena, Italy}
M.~Calvetti, G.~Collazuol,
G.~Graziani, E.~Iacopini, M.~Lenti, A.~Michetti \\
{\small \em Dipartimento di Fisica dell'Universit\`a e Sezione
    dell'INFN di Firenze, I-50125 Firenze, Italy} \\[0.35cm]
%
H.G.~Becker,
P.~Buchholz,\footnote{Present address: Institut f\"ur Physik, Universit\"at
           Dortmund. D-44221 Dortmund, Germany}
D.~Coward,$^{9)}$
C.~Ebersberger, H.~Fox, A.~Kalter, K.~Kleinknecht, U.~Koch, L.~K\"opke,
B.~Renk, J.~Scheidt, J.~Schmidt, 
V.~Sch\"onharting, Y.~Schu\'e,
R.~Wilhelm, M.~Wittgen \\
{\small \em Institut f\"ur Physik, Universit\"at Mainz, D-55099
    Mainz, Germany\/}\footnote{Funded by the German Federal Minister for
    Research and Technology (BMBF) under contract 7MZ18P(4)-TP2} \\[0.35cm]
J.C.~Chollet, S.~Cr\'ep\'e,  L.~Fayard, L.~Iconomidou-Fayard, 
J.~Ocariz,\footnote{Permanent address: Departamento de F\'{\i}sica,
Universidad de los Andes, M\'erida 5101-A, Venezuela}
G.~Unal, D.~Vattolo, I.~Wingerter-Seez \\
{\small \em Laboratoire de l'Acc\'{e}l\'{e}rateur Lin\'{e}aire, IN2P3-CNRS,
Universit\'{e} de Paris-Sud, \\
F-91406 Orsay, France\/}\footnote{Funded by Institut National de Physique
des Particules et de Physique Nucl\'{e}aire (IN2P3), France} \\[0.35cm]
G.~Anzivino, F.~Bordacchini, P.~Cenci, P.~Lubrano, A.~Nappi, 
M.~Pepe, M.~Punturo\\
{\small \em Dipartimento di Fisica dell'Universit\`a e Sezione
    dell'INFN di Perugia, I-06100 Perugia, Italy} \\[0.35cm]
L.~Bertanza, A.~Bigi,
P.~Calafiura,\footnote{Present address: E.O. Lawrence Berkeley 
National Laboratory, Berkeley, CA 94720 USA} 
R.~Carosi, R.~Casali, C.~Cerri, M.~Cirilli,
F.~Costantini, R.~Fantechi, S.~Giudici, 
B.~Gorini,$^{1)}$
I.~Mannelli,  V.~Marzulli, 
G.~Pierazzini,
F.~Raffaelli, M.~Sozzi \\
{\small \em Dipartimento di Fisica dell'Universit\`a,
Scuola Normale Superiore\\
e Sezione dell'INFN di Pisa, I-56100 Pisa, Italy} \\[0.35cm]
J.B.~Cheze, M.~De Beer, P.~Debu,  
R.~Granier de Cassagnac, 
P.~Hristov,\footnote{Present address:  Joint Institute for Nuclear Research, Dubna, Russian
Federation}
E.~Mazzucato, S.~Schanne,  R.~Turlay, B.~Vallage \\
{\small \em DSM/DAPNIA - CEA Saclay, F-91191 Gif-sur-Yvette,
    France} \\[0.35cm]
I.~Augustin,$^{1)}$
M.~Bender, M.~Holder, M.~Ziolkowski \\
{\small \em Fachbereich Physik, Universit\"at Siegen, D-57068 
Siegen, Germany\/}\footnote{Funded by the German Federal Minister for
Research and Technology (BMBF) under contract 056SI74} \\[0.35cm]
R.~Arcidiacono, C.~Biino,  R.~Cester, F.~Marchetto, E.~Menichetti,
N.~Pastrone \\
{\small \em Dipartimento di Fisica Sperimentale dell'Universit\`a e
    Sezione dell'INFN di Torino, \\ I-10125 Torino, Italy} \\[0.35cm]
J.~Nassalski, E.~Rondio, M.~Szleper, W.~Wislicki, S.~Wronka \\
{\small \em Soltan Institute for Nuclear Studies, Laboratory for High
    Energy Physics, \\ PL-00-681 Warsaw, Poland\/}\footnote{
   Supported by the Committee for Scientific Research grant 2P03B07615
    and using computing resources of the Interdisciplinary Center for
    Mathematical and 
    Computational Modelling of the University of Warsaw} \\[0.35cm]
H.~Dibon, G.~Fischer, M.~Jeitler, M.~Markytan, I.~Mikulec, G.~Neuhofer,
M.~Pernicka, A.~Taurok \\
{\small \em \"Osterreichische Akademie der Wissenschaften, Institut
    f\"ur Hochenergiephysik, \\ A-1050 Wien, Austria\/}\footnote{
    Funded by the Austrian Ministery for Traffic and Research under the
   contract
   GZ 616.360/2-IV GZ 616.363/2-VIII, Austria and by the Fonds f\"ur
    Wissenschaft und Forschung FWF Nr. P08929-PHY} \\[0.35cm]
}
\end{center}
\thispagestyle{empty}

\newpage
\pagenumbering{arabic}  
\section{Introduction}
CP violation has been observed so far only in the decays of neutral
kaons~\cite{cp}.  The main component of the effect~\cite{icp} occurs in
the mixing between $\ko$ and $\kobar$ eigenstates.  The physical
states $K_S$ and $K_L$ deviate from pure CP=$\pm 1$ eigenstates, with the
mixing described by the parameter $\varepsilon$. On the other hand, direct
CP violation can occur in the transition from the 
neutral kaon eigenstate with CP=$-1$ 
to a two--pion final state, which has CP=$+1$. 
This can be observed through the interference of the amplitudes of different
final state isospin~\cite{dcp_pipi} and is described by the parameter
$\varepsilon^\prime$. The measured quantity is the 
double ratio of decay widths:
\begin{equation}
 R \; = \; 
   \frac{\Gamma(\kl \rightarrow \,\pipin\,) /\, \Gamma(\ks \rightarrow \,\pipin\,)}
        {\Gamma(\kl \rightarrow \pipic) /\, \Gamma(\ks \rightarrow \pipic)} 
      \simeq 1-6\times \repr.
\end{equation}

The Standard Model relates CP violation to the existence of three
generations of quarks and to the complex phase present in the CKM
matrix~\cite{dcp}.
Computations based on the Standard Model typically predict $\repr$ in
the range 0 to $10\times 10^{-4}$~\cite{eprime_th}.
The first evidence for direct CP violation was published in 
1988~\cite{na31first}.
Subsequent measurements $(23.0\pm 6.5) \times 10^{-4}$~\cite{na31}
and $(7.4 \pm 5.9) \times 10^{-4}$~\cite{e731} 
have been only marginally consistent with each other; 
the measurement in~\cite{na31} shows evidence for
direct CP violation while that in~\cite{e731} is consistent with no
effect.  The KTeV collaboration has recently measured $(28.0 \pm 4.1)
\times 10^{-4}$~\cite{katev} using a technique similar to the one used
in~\cite{e731}.

In this letter, a new measurement of $\repr$ obtained by
the NA48 collaboration at CERN is presented, 
based on the first data collected with
the experiment in 1997.  
While the ultimate accuracy of the NA48
experiment will be significantly better than that reported in this study,
this result is of sufficient precision to help clarify the current
experimental situation.

\section{Principle of measurement}
The measurement of $\repr$ proceeds by counting the number of events
in each of the four decay modes and computing the double ratio $R$.
The NA48 experiment is designed in such a way that many systematic
uncertainties can be controlled with high accuracy 
by exploiting cancellations of effects which contribute
symmetrically to different components of the double ratio.

Data are collected simultaneously in the four decay channels,
minimizing the sensitivity of the measurement to accidental activity
and to variations in beam intensity and detection efficiency.  $\kl$
and $\ks$ decays are provided by two nearly collinear beams, with
similar momentum spectra, converging in the centre of the main
detector.

In order to reduce the difference in acceptance due to the large
difference in average decay lengths, only $K_L$ decays occurring in
the region also populated by $K_S$ decays are used for the measurement
of the double ratio.  Furthermore, a cancellation of the residual
difference is obtained by weighting the $K_L$ events used in the double
ratio with a function of the proper lifetime $\tau$, proportional to
the expected ratio of $\ks$ and $\kl$ decays at time $\tau$.  In this
way, the systematic accuracy of the result does not rely
on a detailed Monte Carlo simulation of the experiment.

\section {Beam and detector}

\subsection {The kaon beams}

The neutral beams~\cite{beam} are derived from 450 GeV/$c$ protons
extracted from the CERN SPS.  Due to the different mean decay lengths
($\lambda_L$=3.4~km, $\lambda_S$=5.9~m at the average kaon momentum
of 110~GeV/$c$), two separate production targets are used, located
126~m and 6~m upstream of the beginning of the decay region.

For each SPS pulse (2.4~s spill every 14.4~s), $1.1\times 10^{12}$
protons hit the $\kl$ production target.  A neutral secondary beam,
with $\pm\, 0.15$~mrad divergence is generated at a production angle of
2.4~mrad. Three stages of collimation are used with the final element
located just upstream of the beginning of the decay region.

Part of the non-interacting primary protons impinge on a bent silicon
mono--crystal \cite{crystal_1}.  A small fraction of these protons undergo
channelling, and are bent to produce a collimated beam of $3 \times
10^{7}$ protons per pulse
transported to the $\ks$ production target.

The $\ks$ collimator selects secondary neutral particles at 4.2~mrad
production angle, with a divergence of $\pm\, 0.375$~mrad.  The
different production angles for the $\kl$ and $\ks$ beams are chosen
to minimize the difference in decay momentum spectra.
The $\ks$ beam enters the
fiducial decay volume 68~mm above the $\kl$ beam.
The beams converge with an angle of
0.6~mrad and the axes of the two beams cross at the position of the
electromagnetic calorimeter.
The total flux of $K_L$ through the fiducial region is $\approx 1.5
\times 10^7$ per spill,  and the $K_S$ flux is 
$\approx 2 \times 10^2$ per spill.

The protons directed to the $\ks$ target are detected by an array of
scintillation counters which comprise the tagging
detector~\cite{tagger}.  This device is used to tag $\ks$ decays.
The time of any protons and the time of the
event in the main detector are each reconstructed relative to a common
free running clock.  The presence (absence) of a proton reconstructed
in coincidence with the event defines the event as $\ks$ ($\kl$).
The signals are digitized by 960~MHz FADCs.  The proton time
resolution is 120~ps and the double-pulse separation is $4$~ns.

The neutral beam produced at the $\ks$ target traverses an
anti--counter (AKS), formed by a set of three scintillation counters
preceded by an aligned 3~mm thick iridium crystal which enhances photon
conversions (1.8 effective radiation lengths)~\cite{crystal_2}.  This
detector is used to veto $\ks$ decays occurring upstream,
thus providing an accurate definition of the beginning
of the decay region.

The decay region is contained in an evacuated  ($< 3 \times 10^{-5}$~mbar)
90~m long tank with a
0.9~mm (0.003 radiation lengths) thick polyimide (Kevlar) 
composite window at the end.
The window extends from an inner diameter of 18.4~cm to an outer
diameter of 2.3~m.  The neutral beam continues in a 16~cm diameter
evacuated tube to a ($\kl$) beam monitor
and the beam dump, both downstream of all the detector elements.

\subsection {Main detector}

\begin {figure}[ht]
\begin{center}
\includegraphics*[scale=0.45]{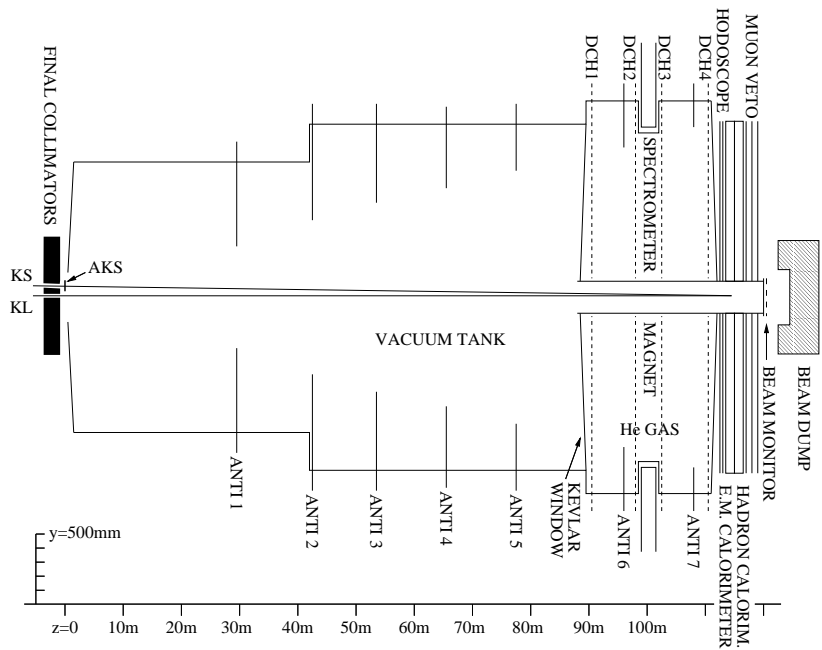}
\caption {Layout of the main detector components.}
\label {fig:detector.eps}
\end{center}
\end {figure}

The layout of the main detector is shown in
figure~\ref{fig:detector.eps}. Kaons which decay into charged pions are
measured by a magnetic spectrometer comprised of four drift
chambers~\cite{dch,dch2}, and a dipole magnet giving a
horizontal transverse momentum kick of 265~MeV/$c$.
The volume
between the drift chambers is filled with helium at atmospheric
pressure.
The drift chambers are each comprised of eight planes of
sense wires, two horizontal, two vertical  and two along
each of the 45$^\circ$ directions.  Only horizontal and
vertical planes are instrumented in DCH3. 
The resolution in each projection is 90~$\mu $m and
the average efficiency is 99.5~\% per plane. The geometrical accuracy
is better than 0.1~mm/m.  The momentum resolution is
$\sigma_p/p$ $\simeq$ 
$0.5\: \% \; \oplus \; 0.009\: \% \times p$ 
($p$ in GeV/$c$), where $\oplus$ means
that the contributions should be added in quadrature.

Two plastic scintillator hodoscope planes are placed after the helium
tank.  They are segmented in horizontal and vertical strips, respectively,
with widths varying between 6.5~cm and 10~cm, arranged in 4 quadrants.
Fast logic  provides a
signal for the first level of the $\pipic$ trigger.  This trigger
requires the presence of tracks in two opposite quadrants of the
hodoscope.  The hodoscope data are also used in the analysis to
determine the event time of $\pipic$ events for the tagging procedure.
The time resolution is $\simeq 200$~ps per track.

A quasi--homogeneous liquid krypton electromagnetic (e.m.) calorimeter
with 127~cm (27 radiation lengths)
long projective tower readout is used to measure the
photons from $\pipin$ events.  The electrode structure consists of
copper/beryllium ribbons extending between the front and the back of
the detector with a $\pm\, 48$~mrad accordion geometry.  The 13212
readout cells each have a cross section of $2 \times 2$~cm$^2$ at the
back of the active region.  The projective towers point to the average
$K_S$ decay position (110~m upstream of the detector) so that the
measurement of the photon positions is insensitive to the
initial conversion depth.  The transverse scale at the back plane of
the calorimeter is measured to better than 0.2~mm/m.  The initial
current induced on the electrodes by the drift of the ionization is
measured using pulse shapers with 80~ns FWHM and digitized with 40~MHz
FADCs~\cite{cpd}.  The energy resolution is $\sigma (E) /E$ $\simeq$
$0.125/E \oplus 0.032/\sqrt{E} \oplus 0.005$  ($E$ in GeV).
The spatial resolution is better than 1.3~mm and the time resolution
is better than 300~ps for 20~GeV photons.

The e.m.~calorimeter incorporates a detector made of a
4~mm thick plane of scintillating fibres located near the maximum of the shower
development.  This device provides an independent trigger for $\pipin$
decays, and an additional measurement of the event time.

An iron--scintillator hadron calorimeter, 6.7 nuclear interaction
lengths thick, is used to complement the krypton calorimeter to
measure the energies of hadrons in the trigger.  Three planes of
scintillation counters, segmented into 25~cm wide counters, shielded
by 80~cm thick iron walls, are used for muon identification to reduce
the background from $\kl \rightarrow \pi\mu\nu$ ($\kmuthree$) events.
Seven counter arrays (indicated as ANTI~1 --
ANTI~7 on figure~\ref{fig:detector.eps}), consisting of
plastic scintillators and iron converters,
surround the vacuum and helium tanks to detect photons which
miss the calorimeter.
In the present analysis, they are only used to
monitor accidental activity.

\subsection {Trigger and DAQ}

The readout of the e.m.~calorimeter sends the calibrated analogue sums
of $2\times 8$ cells to the $\pipin$ trigger.  The
$\pipin$ trigger~\cite{nut} digitizes the data and operates as a 40~MHz
pipelined computation system with a latency of 3~$\mu$s.  The
calorimeter data are reduced to $x$ and $y$ projections, each
consisting of 64 strips, which are used to reconstruct the total deposited
energy $E_0$, the first and second moments of the energy distribution
and to reconstruct and measure the time of energy peaks.  This
information is used to reconstruct the radial position of the 
centre of gravity $C$ of the
event and the proper decay time $\tau/\tau_S$ in units of the mean
$\ks$ lifetime $\tau_S$.  Events are selected if $E_0 > 50$~GeV, $C
< 15$~cm, $\tau/\tau_S < 5.5$ and if there are no more than five
peaks in both views in a 9~ns time window.  The $\pipin$ trigger rate
is about 4500 per pulse and the deadtime is negligible.

A two level $\pipic$ trigger is used.  At the first level, the
hodoscope trigger is placed in coincidence with a total energy
condition ($\ge 30$~GeV), defined by adding the hadron calorimeter energy
to $E_0$ from the $\pipin$ trigger.  The first level trigger is
downscaled by a factor of two, the resulting trigger rate is about
100,000 per pulse.  The second level $\pipic$ trigger~\cite{mbox} uses
information from the drift chambers.  It consists of hardware
coordinate builders for each view and a farm of asynchronous
microprocessors which reconstructs tracks.  
Triggers are selected if the tracks converge to within 5~cm, their
opening angle is less than 15~mrad, the reconstructed proper decay time
is less than 4.5 $\tau_S$ 
and the reconstructed kaon mass is at least 95~\% of the
nominal $\ko$ mass.  The maximum decision latency is 100~$\mu$s. The
second level trigger rate is about 2000 per pulse.

Downscaled triggers,
based on relaxed conditions are collected to compute the efficiency of
the main triggers.  ``Random events'' record any activity in the
detectors 69~$\mu$s after a heavily downscaled beam counter signal.
The 69~$\mu$s delay corresponds to the periodicity of the slow proton
extraction (three revolution periods) of the SPS accelerator.  These
events contain signals which are representative of accidental activity
in the detectors.

The overall deadtime of the trigger system is below 1~\%,
the largest contribution being due to the $\pipic$ trigger.  The
deadtime condition is recorded and made completely symmetric offline.
The detector readout systems provide information on all the activity
in the detectors in a readout window of about 250~ns surrounding the
trigger time.  The readout window for the drift chambers is
larger, about 1.2~$\mu$s.

\section{Event reconstruction and selection}

\subsection{\boldmath{$\pipin$} mode}

Photon showers are found in the calorimeter by looking for maxima in
the digitized pulses from individual cells in both space and time and
accumulating the energy within a radius of 11~cm. Photon energies
are corrected for: energy outside the cluster
boundary; energy lost in non-working cells (about $0.4\,\%$
of the channels); energy from the
other photons in the event; small variations of the energy
measurement depending on the impact point within the cell and 
small space charge effects from the accumulation of positive
ions~\cite{spacecharge} because the calorimeter was operated at half
the nominal $3\,$kV drift voltage in 1997.

$K \rightarrow$ $\pipin$ decays are selected by requiring four clusters
with energies between 3~GeV and 100~GeV which are in time within $\pm\,
5$~ns of the average of the four.  Fiducial cuts are applied to ensure
that photon energies are well measured.  These cuts include the
removal of photons within 2~cm of a non-working cell
and photons within a $\pm\,
4$~cm wide column due to a failure in a HV connection during the 1997
run.  The minimum distance between
photon candidates is required to be greater than 10~cm.  To reduce the
background from $K_L \rightarrow 3\,\pi^0$ decays, events with an
additional cluster of energy above 1.5 GeV and within $\pm\, 3$~ns of
the $\pipin$ candidate are rejected.
 
The kaon energy is obtained with a resolution of $\sim 0.6$~\% from
the sum of the photon energies.  The longitudinal position of the
decay vertex relative to the front of the calorimeter $D$ is
reconstructed from the energies $E_i$ and positions $x_i, y_i$ of the
four selected clusters, assuming that their invariant mass is the kaon
mass ($m_K$) as follows:
\begin{equation}
 D = \sqrt{\sum_i \sum_{j>i}E_iE_j[(x_i-x_j)^2+(y_i-y_j)^2]}\: /\: m_K
\end{equation}
and is used to reconstruct the proper decay time $\tau$.  The resolution
of $D$ is $50$ to $70$~cm depending on energy.  The invariant masses 
$m_1$ and $m_2$ of the two photon pairs are 
computed using $D$ and
compared to the nominal $\pi^0$ mass ($m_{\pi^0}$).  For this, a
$\chi^2$ variable is computed as follows:
\begin{equation}
 \chi^2 = 
\left[ \frac{(m_1+m_2)/2-m_{\pi^0}}{\sigma _+} \right ]^2 \; + \;  
\left[ \frac{(m_1-m_2)/2}{\sigma_-} \right ]^2
\end{equation}
where $\sigma_\pm$ are the resolutions of $(m_1\pm m_2)/2$ observed in
the data (parameterised as a function of the lowest photon energy).
The three-fold choice of photon pairings is made by selecting the
combination with the lowest $\chi^2$.  Due to the constraint on $m_K$,
$(m_1 \pm m_2)$ are to good approximation uncorrelated.  Typical
values of $\sigma_+$ and $\sigma_-$ are 0.45 and 1.1~MeV$/c^2$.  To select
good $\pipin$ candidates, the best combination is required to have
$\chi^2 < 13.5$. This requirement removes about 7~\% of $K_S
\rightarrow \pipin$ events in a pure $K_S$ beam where there is no
background.  About 70~\% of these are due to $\pi^0$ Dalitz decays and
to photon conversions in the Kevlar window or in the spectrometer
volume upstream of the magnet. The remaining part is due to the
e.m.~calorimeter response, and is dominated by occasional photon--nucleon
interactions, which result in a reduction of the observed energy.
Both effects are highly symmetric between $K_S$ and $K_L$, and
do not affect the measurement of the double ratio.

The event time as used in the tagging procedure for $\pipin$ events is
computed combining eight time estimators from the two most energetic
cells of each cluster.

\subsection{\boldmath{$\pipic$} mode}

Tracks are reconstructed from the hits and drift times in the
spectrometer.  A vertex is defined at the point of closest approach of
two oppositely charged tracks (provided they approach to within 3~cm).
The momenta of the tracks are derived using a detailed magnetic field map.
The energy of the event is reconstructed using
the angle between the tracks before the magnet and the ratio of the
track momenta, assuming the two tracks originate from $K \rightarrow
\pipic$.  This avoids systematic uncertainties from the 
knowledge of the magnetic field
in the definition of the fiducial energy cut and lifetime measurement.
Small corrections on the tracks are made for the field (integral $\sim
2\times 10^{-3}$~T~m) in the decay region before the spectrometer.

The vertex resolution is typically $50$~cm in the longitudinal
direction $z$, and $2$~mm in the transverse directions $x$, $y$.
Since the beams are separated by about 6~cm in the decay region, a
clean identification of $K_S \rightarrow \pipic$ and $K_L \rightarrow
\pipic$ decays is possible (this is referred to as ``$y$-vertex
tagging'' in the following).

A variable $\asp$ related to the decay orientation in the kaon rest frame
is defined from the two track momenta $p_1$ and $p_2$ as $\asp
= |p_1 - p_2|/(p_1+p_2)$.  Kaon energy dependent cuts $\asp < 0.62$
and $\asp < 1.08-0.0052\times E_K$ (where $E_K$ is 
the kaon energy in GeV) are made in order
to avoid using events with tracks near the beam hole for which a
precise Monte Carlo modelling is required.  This cut also entirely
removes $\Lambda$ and $\bar{\Lambda}$ decays which are present in the
$\ks$ beam.

To reject background from semileptonic $K_L$ decays, tracks consistent
with being either an electron or a muon are rejected.  To identify
electrons, the ratio $E/p$ of the energy of a matching cluster
in the e.m.~calorimeter to the track momentum is computed.
Both tracks must satisfy $E/p < 0.8\,$. This requirement
reduces the $\kl \rightarrow \pi e\nu$ ($\kethree$) 
background by a factor of 500 while removing
about 5~\% of the $\pipic$ signal.  Tracks are identified as muons if
hits are found in time in the muon counters near the extrapolated track
impact point.  Events with identified muons are rejected.  The
rejection against $\kmuthree$ background is about 500 with an
efficiency for $\pipic$ of 97~\% (the loss coming essentially
from $\pi^{\pm}$ decays to $\mu^{\pm}$).

For good $\pipic$ events, the invariant mass $m_{\pi\pi}$ 
should be equal to the kaon mass.
The invariant mass resolution $\sigma_m$ is typically $2.5$~MeV/$c^2$.  An
energy dependent cut at $\pm\, 3\,\sigma_m$ is applied.
A further reduction of background due to semileptonic decays is
obtained using a cut based on the kaon transverse momentum.  To
define a selection which is symmetric between $K_S$ and $K_L$ decays,
a variable $\ptprime$, defined as the
component of the kaon momentum orthogonal to the line joining the
production target (identified with $y$-vertex tagging) and the point
where the kaon trajectory crosses the plane of the first drift chamber,
is used.  The $\ptprime$ resolution is independent of the target from
which the event originated.  To select $\pipic$ candidates, the cut
$\ptprime{}^2 < 2 \times 10^{-4}$~GeV$^2$/$c^2$ is applied.  Event
losses ($\simeq 8\times10^{-4}$) are mainly due to large deviations in
the vertex reconstruction which are independent of whether the event
is $\kl$ or $\ks$.

The event time as used in the tagging procedure in the $\pipic$ mode
is measured by combining the four times from the two hodoscope planes for
each track.  The procedure which is applied protects against tails in
single time measurements (predominantly due to accidental hits).
Events for which this procedure cannot be applied are rejected,
removing $2.3\times10^{-3}$ of the $\pipic$ signal.

\subsection{Tagging}

\begin {figure}[ht]
\begin{center}
\includegraphics*[scale=0.43]{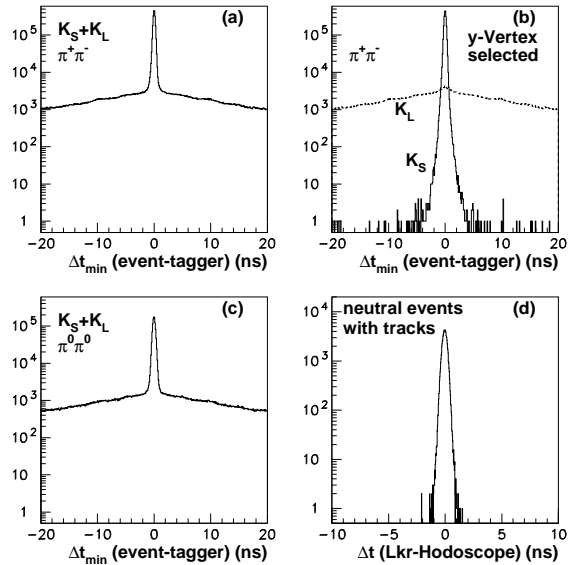}
\caption {(a) Distribution of the minimum difference between
  tagger and event times for $\pipic$ decays.  The peak corresponds to
  $\ks$ events where there is a coincidence. (b) The same
  distribution, separated into $\ks$ and $\kl$ events using $y$-vertex
  tagging.  This shows that the tails in the coincidence are very
  small.  (c) as (a) for $\pipin$ mode.  (d) Coincidence time for
  neutral events with tracks (see text).}
\label {fig:tagging.eps}
\end{center}
\end {figure}

The event is tagged by classifying it as $K_S$ if there is a proton
within a $\pm\, 2$~ns coincidence window centred on the event time,
otherwise the event is classified as $K_L$ as shown in
figure~\ref{fig:tagging.eps}a,c.
 
An accidental coincidence between a proton traversing the tagger and a
$K_L$ event may cause the event to be wrongly counted as  $K_S$.
This probability is measured for $\pipic$ events to be
$\alpha^{+-}_{LS} = 0.1119\pm 0.0003$ using $y$-vertex tagging.  The
measurement of the double ratio is sensitive to the difference in this
probability between $\pipin$ and $\pipic$ events, which is measured
to be $\alpha^{00}_{LS}-\alpha^{+-}_{LS} = (10\pm 5) \times 10^{-4}$
leading to a correction to $R$ of $(18 \pm 9) \times 10^{-4}$.  This
is measured by looking at proton coincidences at times which are
offset from the actual event.
Differences in accidental rate between event time and offset coincidence
intervals are checked with
$\kl \rightarrow \pipic$ events, identified from
y-vertex tagging, and with $\kl \rightarrow 3\pi^0$ events.  The observed
difference in effective rate seen by $\pipin$ and $\pipic$ events is
compatible with the intensity dependence of the $\pipic$ trigger
efficiency.
 
Inefficiencies in reconstructing a proton or errors in measuring the
event time or the proton time can cause a coincidence to be lost and a
true $K_S$ event would be classified as $K_L$.  From the $y$-vertex
tagged $\pipic$ events, the probability for this to happen is
$\alpha^{+-}_{SL} = (1.5\pm 0.1) \times 10^{-4}$
(figure~\ref{fig:tagging.eps}b).  Detailed studies show that this is
dominated by inefficiencies in the tagger which are intrinsically
symmetric between $\pipin$ and $\pipic$ decays.  A sample of neutral
events containing $\pi^0$ Dalitz decays and photon conversions is used
to compare the time measured using the calorimeter with the time of
the tracks measured using the hodoscope
(figure~\ref{fig:tagging.eps}d).  This comparison provides the
difference between tagging efficiencies for $\pipin$ and $\pipic$
events, which is found to be zero, with an accuracy of $\pm\, 1 \times
10^{-4}$, leading to an uncertainty on $R$ of $\pm\, 6 \times 10^{-4}$.
As an independent check, $\pipin$ events containing a Dalitz decay
have been used to $y$-vertex tag $\pipin$ events leading to a direct
measurement of $\alpha_{SL}^{00} = (2.3^{+3.0}_{-1.0})\times 10^{-4}$,
in agreement with the method above.

\subsection{Data quality selections, data samples}

Data collected in the $\pipic$ mode are affected by an overflow
condition in the drift chambers which resets the front-end readout
buffers when there are more than seven hits in a plane within a 100~ns
time interval~\cite{dch2}.  This occurs when an accidental particle
generates an electromagnetic shower upstream of the spectrometer and
sprays the drift chambers with particles.  Overflows in each plane are
recorded for at least 400~ns before and after each event.
To maintain the highest detector efficiency, we
require that events used in this analysis do not have overflows within
$\pm\, 312$~ns of the trigger time. This interval, which is larger
than the sum of the
maximum drift time and reset time, excludes any correlation between
overflows and selected events.  To avoid indirect biases
related to different average beam intensities, the selection on
overflows is also applied to the $\pipin$ mode resulting in a loss of $20$~\%
of the events.

The fiducial range in kaon energy and proper time used to count events is
$70 < E_K < 170$~GeV and 
$0 < \tau < 3.5 \;\tau_S$, where $\tau=0$ is defined 
at the position of the AKS counter. 
For $K_L$, the decay time cut is applied on reconstructed $\tau$,
while for $K_S$ the cut at $\tau=0$ is applied using the AKS to veto
events occurring upstream.
The nominal
$\tau=0$ position differs by $21.0 \pm 0.5$~mm between $\pipin$ and
$\pipic$ modes, which is accounted for in the analysis.
The veto efficiency of the AKS is $0.9964\pm 0.0003$
($0.9923\pm 0.0010$) for $\pipin$ ($\pipic$) decays and leads to a
negligible correction to $R$.

All events are required to satisfy a symmetric cut on the position of
the centre of gravity.  For $\pipin$ decays, the centre of gravity is
defined as the energy--weighted average $x$, $y$ position of the four
showers at the face of the e.m.\ calorimeter.  For $\pipic$ decays it
is the momentum--weighted average position of the track segments,
measured upstream of the spectrometer magnet and projected onto the
face of the e.m.~calorimeter. The radial position of
the centre of gravity is required to be within 10~cm of the beam axes.
The sensitivity to differences in
beam halo and to $K_S$ scattering in the collimator and in the AKS
counter is minimized by applying the same cut to both modes.
For scattered $K_S$ events, this cut is stronger than the
$\ptprime$ cut applied in the selection of charged events.
It is however relatively wide compared to both the $\kl$ and $\ks$ beam--spot
radii, which are 3.6~cm and 4.6~cm respectively, so that effects related to
resolution smearing can be neglected.

Table~\ref{table:samples} shows the number of events collected in each 
channel after removing background and correcting for 
mistagged events.  The numbers are shown here without lifetime weighting. 

\begin{table}[ht]
\caption{Statistical samples (in thousands of events).}
\label{table:samples}
\center
\begin{tabular}{|c|c|c|c|}

\hline
$K_L \rightarrow \pi^0\pi^0$ & 
$K_S \rightarrow \pi^0\pi^0$ &
$K_L \rightarrow \pi^+\pi^-$ &
$K_S \rightarrow \pi^+\pi^-$ \\
\hline
489 & 975 & 1,071 & 2,087 \\
\hline 
\end{tabular}
\end{table}

\section {Computation of $R$ and systematic uncertainties} 

In order to be insensitive to the difference between the beam spectra
(about $\pm\, 10$~\% in the range 70--170 GeV), the computation of $R$
is done in 20 energy bins.  The numbers of $\ks$ and weighted $\kl$
candidates are corrected for the tagging accidentals and
inefficiencies as described above. Corrections for trigger efficiency,
background subtraction, and residual acceptance difference between
$\ks$ and $\kl$ are applied separately in each energy bin.  The
average of $R$ over the bins is done with an unbiased estimator of
$\log(R)$.

\subsection{Trigger efficiency}

The efficiency of the $\pipin$ trigger is determined from events
triggered by the scintillating fibre detector in the e.m.\
calorimeter.  The average $\pipin$ trigger efficiency is measured to be
$0.9988 \pm 0.0004$ and is the same for $\ks$ and $\kl$.

The efficiency of the $\pipic$ trigger is measured on a selected
sample of events from an auxiliary trigger which reproduces the first
level logic. The combined efficiency of the first level coincidence
and the second level is $0.9168 \pm 0.0009$.  The major contribution
to the inefficiency comes from a timing misalignment in the trigger logic
which existed for part of the run.
Additional inefficiencies in the hodoscope trigger amount
to $(68\pm 8)\times 10^{-4}$.
The double ratio is corrected by $(+\,9 \pm 23)\times 10^{-4}$ for the
difference between the $\pipic$ trigger efficiency for $K_S$ and
$\tau$-weighted $K_L$ decays.

\subsection{\boldmath{$\pipin$} background}
The background in the $\kl \rightarrow \pipin$ sample originates
from $\kl \rightarrow 3\pi^0$ decays with two undetected
photons. 

\begin {figure}[ht]
\begin{center}
\includegraphics*[scale=0.45]{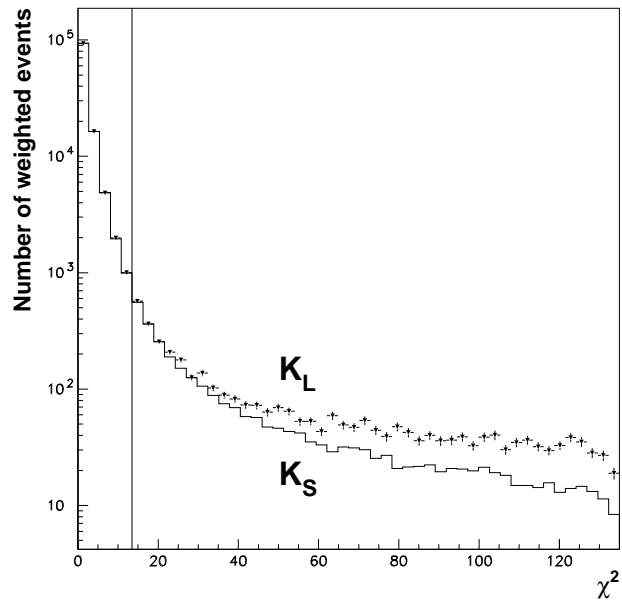}
\caption{Distribution of $\chi^2$ for weighted $K_L$ candidates
(black triangles) compared to the shape for pure 2$\pi^0$ events
derived from $K_S$ candidates, normalized in the first bin. The excess
of events in the $K_L$ candidates in the region
$36$ $<$ $\chi^2$ $<$ $135$ is used to extrapolate the background in
the signal region ($\chi^2$ $<$ 13.5).}
\label {fig:bkn_relli.eps}
\end{center}
\end {figure}

The estimation of the residual background after all cuts is done by
extrapolating the number of events in the $\chi^2$ control region
36--135 to the signal region $\chi^2 < 13.5$ as shown in
figure~\ref{fig:bkn_relli.eps}. The extrapolation factor for the
background is measured using a sample of $\kl \rightarrow 3\pi^0$
simulated events and is found to be $1.2\pm0.2$ compared to a flat
extrapolation (where the error is due to Monte Carlo statistics).

Due to non-Gaussian tails in the $\chi^2$ distribution, the control
region contains a mixture of $\kl \rightarrow 3\pi^0$ and $\pipin$
events.  The number of $\pipin$ events is estimated and subtracted by
using the $\chi^2$ distribution shape from tagged $\ks$ events (after
correcting for the contamination of accidentally mistagged $\kl$) and
normalising this shape to the $\kl$ $\chi^2$ distribution at small $\chi^2$.

Using this method, the background to the weighted $K_L$ sample is
found to be $(8\pm2) \times 10^{-4}$ averaged over energy, where the
error comes mostly from the Monte Carlo statistical error on the
extrapolation factor.  Changes in the choice of the control region
lead to negligible variations in the estimated background. This
procedure is done separately in each energy bin for the R computation,
and the background is found to increase from $\approx$ 0 at 70 GeV to
$\approx$ 0.2~\% at 170 GeV.

\subsection{\boldmath{$\pipic$} background}

\begin {figure}[ht]
\begin{center}
\includegraphics*[scale=0.45]{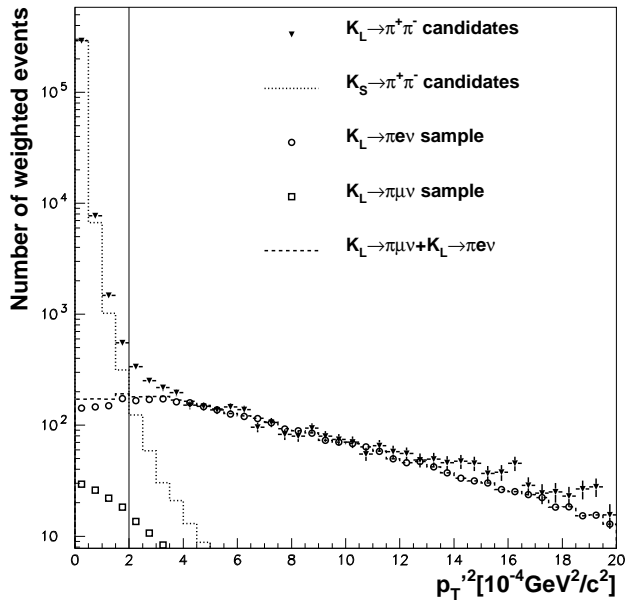}
\caption{Distribution of $p_t{}'{}^2$ for $K_L$ candidates fulfilling
the $\pi^+\pi^-$ invariant mass cut and result
of the fit for the background
determination in the control region, including a signal component (derived from $K_S
$ events),
a $\kethree$ component and a $\kmuthree$ component.
The final signal region
is $p_t{}'{}^2$ $<$ $2\times 10^{-4}$ GeV$^2$.}
\label {fig:chbkg_pl_1d.eps}
\end{center}
\end {figure}

Semileptonic decays of $\kl$ are selected by identifying the lepton as
an electron (from $E/p \simeq 1$) or a muon (from associated hits in
the muon counters and low $E/p$).  For each kaon energy bin the
event distribution in the $(m_{\pi\pi}, p_t{}^\prime{}^2)$ plane is
determined.  The sample of $K_L\rightarrow \pi^+\pi^-$ is then fitted
as the sum of the contribution from true $\pi^+\pi^-$ decays
(determined from the shape of $K_S\rightarrow \pi^+\pi^-$), and from
the $\kethree$ and $\kmuthree$ distributions.  Systematic
uncertainties are evaluated determining the background in different
control regions (contained in the
domain $475 \leq m_{\pi\pi} \leq 505$~MeV/$c^2$
and $p_t{}^\prime{}^2 \leq 2\times10^{-3}$~GeV$^2$/$c^2$) and
verifying the stability of the result for different sub-samples of
data.  The background in the signal region is $(23 \pm 2
\mathrm{(stat)} \pm 4 \mathrm{(syst)} )\times 10^{-4}$, where the
result is averaged over the kaon energy spectrum.  The larger
contribution is from $\kethree$ decays.
Figure~\ref{fig:chbkg_pl_1d.eps} shows the event distribution in the
variable $p_t{}^\prime{}^2$, for $K_L$ and $K_S$ events, together with
the result of the fit to the background.

Kaon decays to $\pipic\gamma$ have been investigated and
their effect on $R$ is negligible.

A small fraction ($\simeq 0.5 \times 10^{-3}$) of $K\rightarrow
\pi^+\pi^-$ are found in the $\kl$ beam at large values of
$p_t{}^\prime{}^2$, extending beyond $(3$--$5)\times
10^{-3}$~GeV$^2$/$c^2$.  These are due to scattering
and regeneration on the collimators.  Since these
events are removed from the $\pipic$ sample by the $p_t{}^\prime{}^2$
cut, but satisfy the centre of gravity cut 
and are kept in the $\pipin$ sample, they
cause a correction to the observed value of the double ratio equal to
$(-12 \pm 3) \times 10^{-4}$.

\subsection{Energy and distance scale calibrations, stability, linearity}

The measurements of the kaon energy, the decay vertex and the proper
time in the $\pipin$ mode rely on the measurement of the photon
energies and positions in the calorimeter.  The calorimeter
performances are studied and adjusted using the $\pipin$ events
themselves, $\kethree$ decays where the electron energy measured in
the calorimeter can be compared to the momentum measured in the
spectrometer, and a so-called $\eta$ run which was a special run
with a $\pi^-$ beam striking a thin target near the AKS counter and
producing $\pi^0$ and $\eta$ ($\rightarrow \gamma\gamma$, 3$\pi^0$)
with known decay position.

\begin {figure}[ht]
\begin{center}
\includegraphics*[scale=0.45]{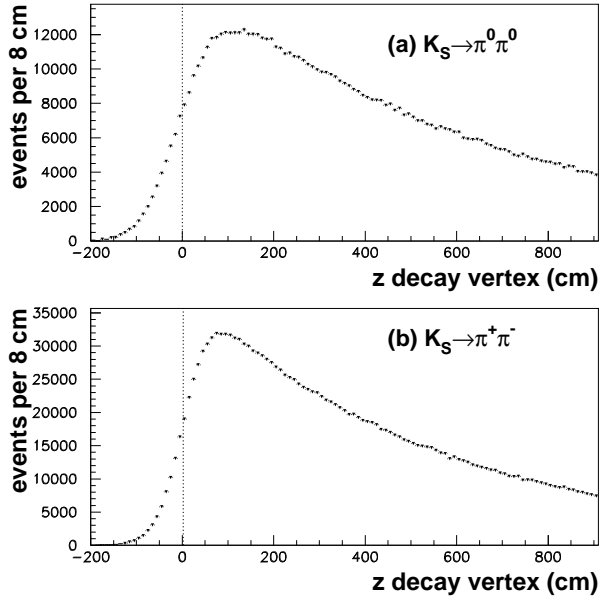}
\caption{Distribution of the reconstructed decay vertex in $K_S$ events
for (a) the $\pipin$ mode and (b) the $\pipic$ mode. The
rising edge corresponds to the position of the anticounter
(represented by the dotted line).}
\label{fig:aks_decay.eps}
\end{center}
\end {figure}

The overall energy and distance scale is adjusted using $\ks
\rightarrow \pipin$ events. The beginning of the decay region is
defined by using the AKS as veto.  Therefore the distribution of the
reconstructed vertex position exhibits a sharp rising edge at the
nominal anti-counter position, as shown in
figure~\ref{fig:aks_decay.eps}a. The energy scale is set such that the
average value of the reconstructed decay position in a range centred
around the anti-counter matches the value found in a Monte Carlo with
the nominal scale. This procedure takes into account non-Gaussian
tails in the energy resolution.  The energy scale was checked daily and
was stable within $\pm\,5\times10^{-4}$ during the entire run.
This measurement of the energy scale
is checked using the data from the $\eta$ run.  Another check is
done using $\pipin$ Dalitz decays and comparing the neutral vertex
position with the vertex reconstructed using the charged tracks.  The
residual uncertainty on the overall energy scale is estimated to be
$\pm\,5\times10^{-4}$ and this leads to an uncertainty of
$\pm\,5\times10^{-4}$ on the double ratio.

The procedure above assumes that the transverse size of the
calorimeter is known.  The size of the calorimeter is
checked using $\kethree$ decays, comparing the reconstructed cluster
position with the electron track impact point extrapolated to the
calorimeter.  This comparison is performed with an accuracy of
0.3~mm/m on the transverse scale, which leads to an uncertainty on the
double ratio of $\pm\,3\times10^{-4}$.

The uniformity of the calorimeter response over it's surface is
checked and optimised using $\kethree$ decays and $\pi^0$ decays from
the $\eta$ runs.  Residual non-uniformities in the calorimeter
response over large distances are at the level of $\approx$ $10^{-3}$ and
the corresponding systematic uncertainty 
on the double ratio is $\pm\,3\times10^{-4}$.

The calorimeter energy non-linearity is studied using $\kethree$
events, $K_{L,S} \rightarrow \pipin$ events, and $\pi^0$ and $\eta$ decays
from the $\eta$ runs.  Non linearities are found to be at the level of
$0.3$~\% in the energy response between 3 and 100 GeV with possible
energy offsets of $\pm\,20$ MeV. These effects are taken as systematic
uncertainties and lead to an uncertainty on the double ratio of
$\pm\,9\times10^{-4}$.

Other uncertainties in the photon energy corrections
(energy sharing between clusters and space
charge effect), and from residual effects from the difference between
the photon direction and the projectivity direction of the calorimeter
are estimated to be $\pm\,4\times10^{-4}$ on the double ratio.

Adding all the previous effects together in quadrature, the total
systematic uncertainty coming from the measurement of photon energies
and positions is found to be $\pm\,12\times10^{-4}$ on the double ratio.

For charged decays, the vertex is measured from the reconstructed
tracks and is completely determined by the detector geometry.
As a check, the anticounter position is
fitted on the $\ks \rightarrow \pipic$ events 
(figure~\ref{fig:aks_decay.eps}b), and agrees to within 1~cm of the
expected position.
Uncertainties in the geometry and on the residual effect of the
magnetic field in the decay region lead to an 
uncertainty on the double ratio of $\pm\,5\times10^{-4}$.

\subsection{Accidental effects}

The effects of accidental activity in the detector are minimized by
the simultaneous collection of data in the four channels.  Most of the
accidental activity is related to 
$\kl$ decays and to muons originating from the $\kl$ production target.

The accidental rate can be measured directly from the activity in the
detector within the readout window before each event.  For example,
the number of events affected by overflows is the same for $\kl \rightarrow
\pipic$ and $\ks \rightarrow \pipic$ to within the statistical error
of $0.1\,\%$.  The correction to $R$ due to this particular class of
accidentals which prevent the event reconstruction in the drift
chambers is negligible after removing events with overflows
also from the samples of $\pipin$ decays, as discussed in section 4.4.
Other accidentals may however cause
a migration of events over the acceptance boundaries.  This effect
is also expected to be the same for $\ks$ and $\kl$ events due to
their kinematical and geometrical similarity.

A method to measure these effects is to overlay events with
randomly triggered events which represent accidental activity
in the beam, thereby artificially doubling the accidental
intensity. The comparison of original and overlaid events provides a
direct measurement of intensity effects in terms of gains and losses
of events which are at the level of a few percent. As expected the
effects tend to cancel in the double ratio.  The correction due to
accidentals is $(- 2 \pm 14)\times 10^{-4}$, where the error is
dominated by the statistical uncertainty in the sample of the events
used for the overlay analysis.

The overlay technique does not cope with any additional detector
activity in the $\ks$ beam generated by the same proton which produced
the $\ks$ event.  Studies of this background show
that the effect on the double ratio can be neglected.

\subsection{Acceptance correction}

The $\ks$ and $\kl$ acceptances are made very similar in both modes
by using the technique of weighting $\kl$ events according to their
proper decay time so that the effective longitudinal vertex
distribution of $\kl$ events is made equal to the $\ks$ distribution.
The weighting factor includes small contributions from interference
and $\ks$ decays in the $\kl$ beam, both of which are significant only
for kaon energies above 140~GeV.
The residual acceptance differences are studied 
with a full simulation of the beams and detector.
The Monte Carlo sample is five times larger than the event sample.
The main sources of systematic error in the acceptance correction are
due to uncertainties in the positions and divergences of the beams and
a significant component is caused by the dead column in the
e.m.~calorimeter.  The overall $\pipic$ reconstruction efficiency has
been found to be the same for $\ks$ and $\kl$ decays using a Monte
Carlo computation which includes the measured efficiencies of each
drift chamber wire.
The acceptance correction is evaluated and applied in each energy bin.
The overall correction to the double ratio, averaged over kaon
energies in the interval 70---170~GeV is $(29 \pm 11 \mathrm{(MC
  stat)} \pm 6 \mathrm{(syst)})\times 10^{-4}$.

\section{Result}
The effects of the corrections applied to the double ratio are listed
in table~\ref{table:corrections}. 
The systematic errors shown in the first
four lines of table~\ref{table:corrections} are dominated by the
statistics of the control sample used in each study and are therefore
statistical in nature.
\begin{table}[ht]
\caption{Corrections applied to {\em R\/} and systematic uncertainties, in $10^{-4}$ units.}
\label{table:corrections}
\center
\begin{tabular}{|l|r c r|}
\hline 
 Tagging $\Delta\alpha_{LS}$, $\Delta\alpha_{SL}$ 
                                & $+\,18 $ & $\pm$ & $ 11$ \\
 $\pipic$ trigger efficiency    & $+\,9 $  & $\pm$ & $ 23$ \\
 Acceptance                     & $+\,29 $ & $\pm$ & $ 12$ \\
 Accidental effects             & $-\:2 $  & $\pm$ & $ 14$ \\
 $\pipin$ background            & $-\;8 $  & $\pm$ & $ 2$  \\
 $\pipic$ background            & $+\,23 $ & $\pm$ & $ 4$  \\
 Scattering/regeneration        & $-\,12 $ & $\pm$ & $ 3$  \\
 Energy scale and linearity     & $ \; $   & $\pm$ & $ 12$ \\
 Charged vertex                 & $ \,  $  & $\pm$ & $ 5$  \\
\hline
Total correction                & $+\,57 $ & $\pm$ & $ 35$ \\
\hline 
\end{tabular}
\end{table}

\begin {figure}[ht]
\begin{center}
\includegraphics*[scale=0.45]{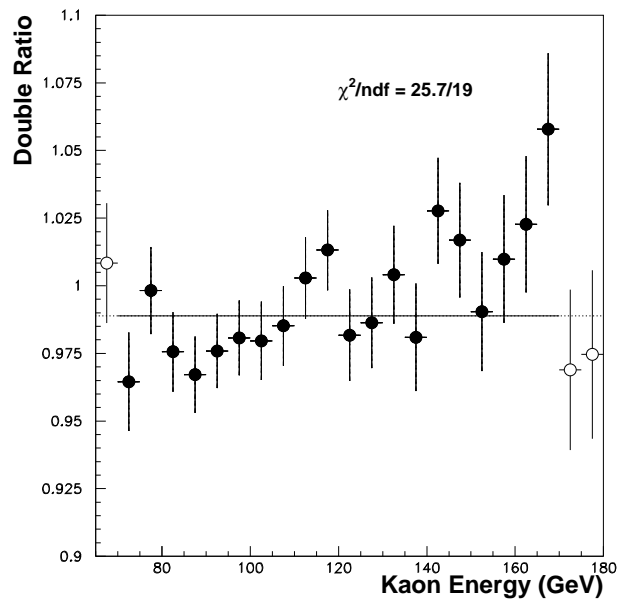}
\caption{Measured double ratio in energy bins. The points used for the
measurement of $\repr$ are shown in black (energy range 70-170 GeV).}
\label {fig:r_e.eps}
\end{center}
\end {figure}

The values of the double ratio in each bin of kaon energy are shown in
figure~\ref{fig:r_e.eps}, after trigger efficiency, tagging,
background and acceptance corrections are included in each bin.  The
bins which are used in the analysis, between 70 and 170~GeV (decided
prior to running the experiment), are shown with black symbols.
The $\chi^2$ of the average is 25.7 (19 d.o.f.).
Extensive checks have been made to exclude systematic
biases which could lead to a variation of $R$ with the kaon
energy as the data might suggest.
As a further check, the double ratio $R$ was computed in three additional
energy bins, as indicated in figure~\ref{fig:r_e.eps}.  These three
extra points strongly disfavour the hypothesis of a linear
trend in the data.

The final result for the double ratio is $R = 0.9889 \pm 0.0027 \pm 0.0035$,
where the first error is statistical and the second is systematic. The 
corresponding value for the direct CP violating parameter is:
$$\mathrm {Re}\,(\varepsilon^\prime\!/\varepsilon) = (18.5 \pm 4.5 \pm
5.8)\times 10^{-4}$$
Combining the two errors in quadrature, the result is:
$$\mathrm {Re}\,(\varepsilon^\prime\!/\varepsilon) = (18.5 \pm 7.3)\times 10^{-4}$$
In conclusion, this new measurement of $\mathrm
{Re}\,(\varepsilon^\prime\!/\varepsilon)$ confirms that direct CP
violation occurs in neutral kaon decays.

\section*{Acknowledgements}
We would like to warmly thank the technical staff of the participating
laboratories and universities for their dedicated effort in the
design, construction, and maintenance of the beam, detector, data
acquisition and processing.

\end{document}